# The Effect of Carbon Nanotube/Organic Semiconductor Interfacial Area on the Performance of Organic Transistors

Narae Kang [1,2,†], Biddut K. Sarker [1,2,†], and Saiful I. Khondaker [1,2,3,*]

[1] NanoScience Technology Center, [2] Department of Physics, [3] Schools of Electrical Engineering and Computer Science, University of Central Florida, Orlando, Florida 32826, USA.

[†] These authors contributed equally to this work
[*] To whom correspondence should be addressed.
E-mail: saiful@ucf.edu

**ABSTRACT**

We show that the performance of pentacene transistors can be significantly improved by maximizing the interfacial area at single walled carbon nanotube (SWCNT)/pentacene. The interfacial areas are varied by anchoring short SWCNTs of different densities (0-30/µm) to the Pd electrodes. The average mobility is increased three, six and nine times for low, medium and high SWCNT densities, respectively, compared to the devices with zero SWCNT. The current on-off ratio and on-current are increased up to 40 times and 20 times with increasing the SWCNT density. We explain the improved device performance using reduced barrier height of SWCNT/pentacene interface.

## 1. Introduction

Organic field-effect transistors (OFETs) have attracted tremendous attention due to their flexibility, transparency, easy processiblity and low cost of fabrication.[1-4] High-performance OFETs are required for their potential applications in the organic electronic devices such as flexible display, integrated circuit, and radiofrequency identification tags.[3,4] A significant research effort has been given in recent years to enhance the performance of the OFETs. Most of the researches were focused to improve the quality of organic semiconductors (OSCs), organic/dielectric interfaces, and other processing parameters.[1,4] One of the major limiting factors in fabricating high-performance OFET is the large interfacial barrier between metal electrodes and OSC which results in low charge injection from the metal electrodes to OSC.[5,6] The interfacial barriers can be caused by several factors such as the discontinuity in morphology, dipole barriers, and Schottky barriers.[7-9] In order to overcome the challenge of low charge injection, carbon nanotubes (CNTs) have been suggested as a promising electrode material for organic electronic devices.[10-15]

Recently, fabrication of OFETs using the CNT electrodes has been reported by several research groups.[10-18] In these reports, the CNT electrodes were fabricated with various techniques using either individual CNT,[10,11] random network CNTs,[15-17] CNT/polymer composite,[12] or aligned array CNTs.[13,14,18] However, an important question remains unanswered: whether the density of CNT in the electrode has any role in the performance of the fabricated OFETs and how much improvement can be achieved using CNT electrode? The density of CNT in the electrodes controls the interfacial area between the CNTs and OSC. A low density CNTs forms small CNT/OSC interfacial area while high density CNTs creates large interfacial area with OSC. It has been suggested from the molecular dynamics simulation and NMR spectroscopy that a π-π interaction exists between CNT/OSC.[19-21] In addition, CNT has a field emission properties due their one-dimensional structure.[22] These theoretical and experimental studies suggest that charge injection should depend on the CNT/OSC interfacial area and that one can improve the performance of OFETs by maximizing CNT/OSC interfacial area. However, no such investigation has been reported yet. Such a study is of great importance for achieving the overreaching goal of the CNT electrodes in organic electronics.

In this paper, we report systematic investigations of the effect of CNT/OSC interfacial area on the performance of the OFETs by varying the density of CNT in the electrode. The devices were fabricated by thermal evaporation of pentacene on the Pd/ single walled CNT (SWCNT) electrodes where SWCNTs of different density (0-30/um) were aligned on Pd using dielectrophoresis (DEP) and cut via oxygen plasma etching to keep the length of nanotube short compared to the channel length. From the electronic transport measurements of 40 devices, we show that the average saturation mobility of the devices increased from 0.02 for zero SWCNT to 0.06, 0.13 and 0.19 $cm^2$/Vs for low (1-5 /μm), medium (10-15 /μm) and high (25-30 /μm) SWCNT density in the electrodes, respectively. The increase is three, six and nine times for low, medium and high density SWCNTs in the electrode compared to the devices that did not contain any SWCNT. In addition, the current on-off ratio and on-current of the devices are increased up to 40 times and 20 times with increasing SWCNT density in the electrodes. Our study shows that although a few nanotubes in the electrode can improve the OFET device performance, significant improvement can be achieved by maximizing SWCNT/OSC interfacial area. The improved OFET performance can be explained due to a reduced barrier height of SWCNT/pentacene interface compared to metal/pentacene interface which provides more efficient charge injection pathways with increased SWCNT/pentacene interfacial area.

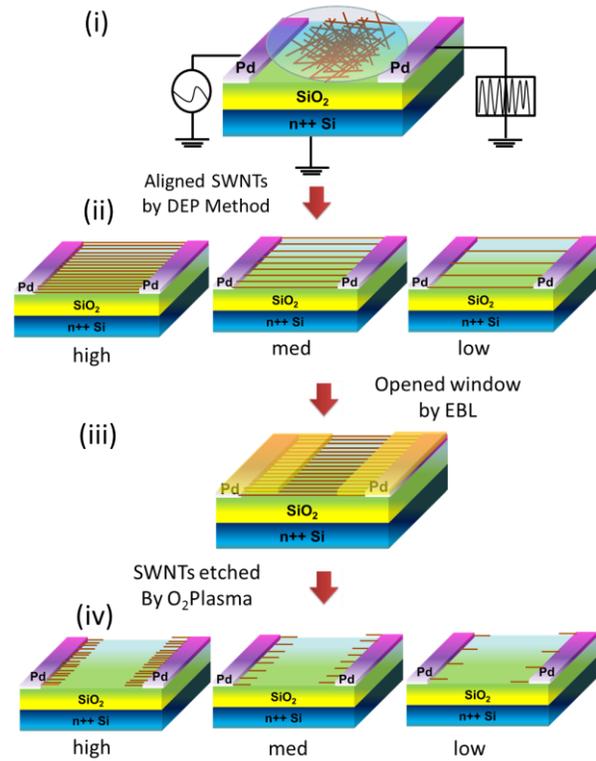

**Figure 1**. Schematic diagram of electrodes fabrication with different SWCNT densities. (i) Assembly of the aligned array SWCNTs by dielectrophoresis (DEP) between the Pd electrodes. (ii) SWNCT assembly with different densities, which were controlled by tuning the SWCNT solution concentration (iii) Opened a window on the SWCNTs array via electron beam lithography and (iv) Etch the SWCNTs by oxygen plasma.

## 2. Experimental details

The devices were fabricated on heavily doped silicon (Si) substrates coated with a thermally grown 250 nm thick silicon di-oxide ($SiO_2$) layer. Palladium (Pd) electrodes of 5 μm x 25 μm were fabricated using standard optical lithography process. The SWCNTs of different linear densities of 0-30/μm were assembled between the Pd electrodes via DEP using a high quality

SWCNT aqueous solution obtained from Brewer Science (Figure 1). The details of the SWCNT assembly can be found in our previous publications.[23] In short, a 3 µl SWCNT solution was dropped onto Pd pattern and an ac voltage of 5 V with a frequency of 2 MHz were applied for 30 sec. Due to the DEP force, the SWCNTs are aligned in arrays between the Pd patterns (Figure 1(i)). The linear density was controlled by varying the concentration of SWCNT solution by diluting the original nanotube solution (~ 50 µg/mL) with deionized (DI) water. The SWCNT arrays were then cut by spin coating PMMA, defining a 4.4 µm ($L$) x 25 µm ($W$) window in the middle of the channel using standard EBL, and subsequent oxygen plasma etching (Figure 1).[24] Finally, the chips are kept into chloroform and cleaned with isopropanol (IPA) and deionized (DI) water. Figure 2(a) shows representative scanning electron microscopy (SEM) images of the part of the electrodes containing an average of 30, 13 and 2 SWCNT/um as well as a bare Pd (zero SWCNT) electrode. The average linear densities of the arrays were calculated by counting the total number of SWCNTs from the SEM images and then dividing it by the channel width. Figure 2(b) shows representative current-voltage ($I$-$V$) characteristics of the arrays before cutting. The typical resistances for the arrays with high, medium and low nanotube density are 0.68 kΩ, 7.19 kΩ, and 63.3 kΩ. As expected, the resistance of the arrays increases with decreasing the density of the SWCNTs in the arrays.[23] Finally, pentacene film with thickness of 30 nm was thermally deposited in vacuum at a pressure of $2\times10^{-6}$ mbar. In order to minimize the device to device fluctuation from the active materials morphology, all of the pentacene films were deposited under identical conditions. The morphological investigation using atomic force microscopy (AFM) showed that all the films have similar morphology with an average grain size of ~150 nm (Figure 3). For a fair comparison of the device performances in terms of nanotube density in the electrodes (different interfacial areas) and to obtain statistically meaningful results, we classified the devices into four categories with a narrow range of SWCNT densities: high (25-30 /µm), medium (10-15 /µm), low (1-5 /µm) and Pd (zero SWCNT) only. The electrical transport measurement of the OFETs were performed using Hewlett-Packed (HP) 4145B semiconductor parametric analyzer connected to a probe station inside an enclosed glove box system with $N_2$ gas flow. A total of 40 devices were investigated with 10 of each category.

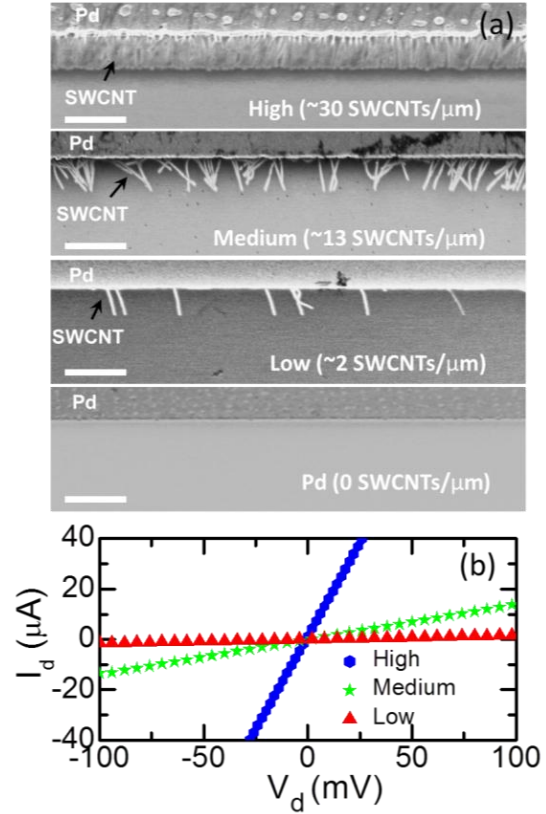

**Figure 2**. (a) SEM images of parts of the source electrodes with high, medium, low density SWCNTs and Pd electrode (scale bar: 500nm) (b) Current-voltage characteristics of the array (before cutting) with with high, medium and low density SWCNTs.

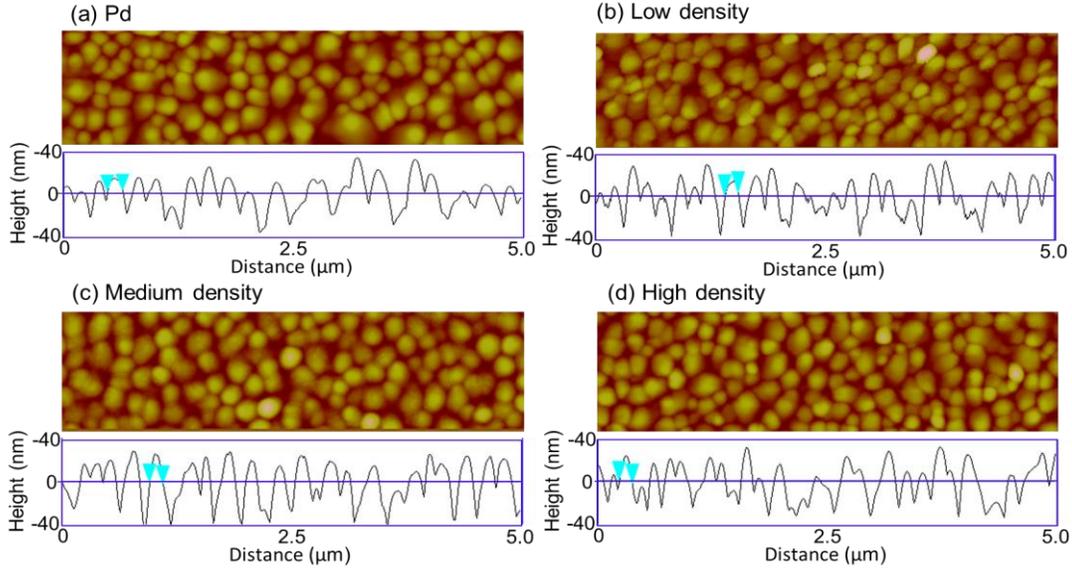

**Figure 3**: Atomic force microscopy (AFM) images of the deposited pentacene film on the electrodes. (a) bare Pd (no SWCNTs), (b) low, (c) medium and (d) high density SWCNT in the electrodes. The height analysis of these films shows the morphology of the films are the similar with typical grain size of ~ 150 nm and rms surface roughness of ~ 3.5 nm

## 3. Results and discussions

Figures 4 (a)-(d) show the drain current ($I_d$) vs source-drain bias voltage ($V_d$) curves (output characteristics) at different gate- voltages ($V_g$) for our best devices with zero, low, medium and high SWCNTs in the electrodes. All the devices show a good gate modulation with linear behavior at low $V_d$ and saturation behavior at higher $V_d$, typical of p-channel OFETs. For comparison of device characteristics, we plotted all the curves in the same scale. From here, we see that the output current significantly increases with increasing the SWCNT density in the electrodes. The output current (at $V_d = -50V$ and $V_g = -20V$) of the devices with zero SWCNTs is 0.15 µA, whereas it is 0.34 µA, 0.81 µA and 1.15 µA for the devices with low, medium and high density SWCNTs in the electrodes. The output current is twice for low density and nine times for the high density SWCNTs compared to the device without any SWCNTs. Since the morphology of all the devices are similar, the increase of output current with increasing SWCNT density clearly show that the interfacial area at the SWCNTs/pentacene has significant impact on the output characteristics of the devices.

To further investigate the effect of the interfacial area on the device performance, we also measured the corresponding transfer curves ($I_d$ vs $V_g$) of the same devices at $V_d = -50V$ (Figure 5 (a)-(d)) and at $V_d = -10V$ and calculated the field effect mobility ($\mu$), on-off ratio ($I_{on}/I_{off}$) and on-current ($I_{on}$) of the devices. The linear mobility $\mu_{lin}$ (at $V_d = -10$ V) and saturation mobility $\mu_{sat}$ (at $V_d = -50$ V) are extracted using the standard formula,[18] $\mu_{lin} = (L/WC_iV_d)(dI_d/dV_g)$ and $\mu_{sat} = (2LI_{d,sat})/(WC_i(V_g-V_T)^2)$, respectively; where $I_{d,sat}$ is saturation current, and $C_i$ is the gate dielectric capacitance (13.8nF/cm$^2$). The maximum $\mu_{sat}$ (maximum $\mu_{lin}$) of the devices for zero, low, medium and high densities SWCNTs in the electrodes are 0.05 (0.03), 0.10 (0.06), 0.19(0.13), 0.29 (0.19) cm$^2$/Vs, respectively. This demonstrates that the mobility of the devices also increases with increasing SWCNT/pentacene interfacial area. The maximum $\mu_{sat}$ is 100%, 280%, and 480% larger for low, medium and high density SWCNTs in the electrode compared to the devices that did not contain any SWCNT. Similar increment in the $\mu_{lin}$ with increasing the SWCNT density is also observed. In calculating the $\mu$, we used $L= 4.4$ µm and $L= 5$ µm for devices with SWCNTs and no SWCNTs respectively. However, the SEM images of Figure 2(a) for low and medium density SWCNTs in the electrode show that there may be an ambiguity in determining $L$ for these densities as the charge injection comes from both Pd and SWCNT interfaces. In order to minimize this uncertainty, we kept lengths of anchored nanotubes to the Pd short (~ 300 nm). Nevertheless, if we were chosen $L= 5$ µm for these two densities then the $\mu_{sat}$ would be 0.11 and 0.22 cm$^2$/Vs, for low and medium SWCNT densities. These values are even higher, and indicate that our experimental data exceeds the error that may arise from the choice of $L$ in low and medium density electrodes. In addition to $\mu$, other important parameters to evaluate the performance of the transistors are $I_{on}/I_{off}$ and $I_{on}$. The transfer curves show that the $I_{on}$ ($I_d$ at $V_g = -80$ V) and $I_{on}/I_{off}$ increase with the SWCNT density in the electrodes. The maximum $I_{on}/I_{off}$ and $I_{on}$ for high density SWCNT electrodes devices are $1.1\times10^5$ and 14.2 µA respectively, whereas they are $3.1\times10^4$ and 12.8 µA for medium density, $1.8\times10^4$ and 10.8 µA for low density, and $9.6\times10^3$ and 3.3 µA for zero density SWCNT in the electrodes. Therefore, both the $I_{on}/I_{off}$ and $I_{on}$ are also increased significantly with increasing SWCNT density in the electrodes.

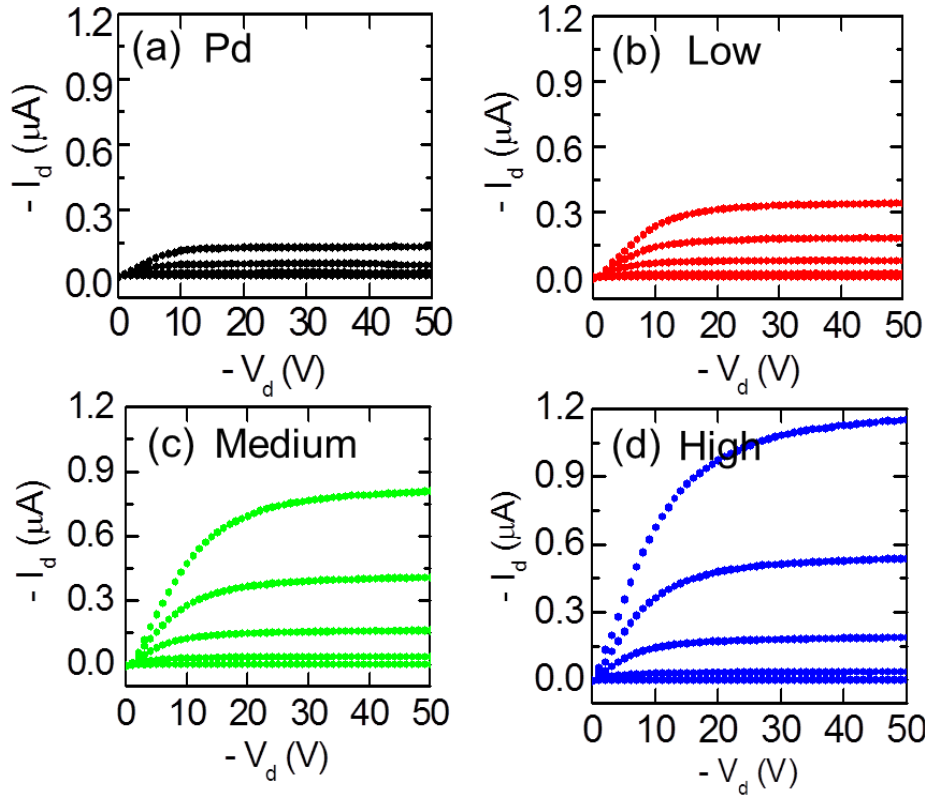

**Figure 4**: Output characteristics ($I_d$-$V_d$) of pentacene transistors at $V_g = 0, -5, -10, -15$ and $-20$ V (bottom to top) for (a) zero, (b) low, (c) medium, and (d) high density SWCNT in the electrodes.

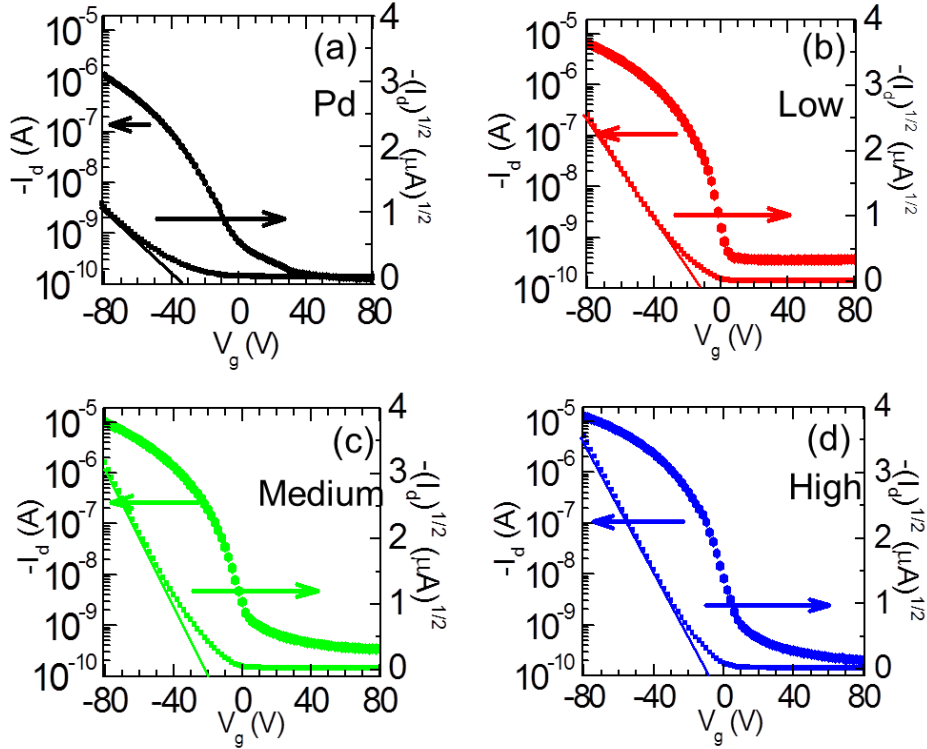

**Figure 5**. Transfer characteristics ($I$-$V_g$ curve) at $V_d = -50$ V (left axis) and $(I_d)^{1/2}$ (right axis) of the devices with (a) zero, (b) low, (c) medium, and (d) high density SWCNT in the electrodes.

The device characteristics measured from 40 devices are summarized in Figure 6 (see also Table 1) where we plot the $\mu$, $I_{on}/I_{off}$ and $I_{on}$ as a function of SWCNT density in the electrodes. Figure 6(a) show that, similar to our best devices, the average $\mu_{sat}$ are increased from 0.02 for zero SWCNT to 0.06, 0.13 and 0.19 cm$^2$/Vs (average $\mu_{lin}$ are increased from 0.01 to 0.03, 0.08 and 0.11 cm$^2$/Vs) for low, medium and high SWCNT density in the electrodes, respectively. The increase in average mobility for our OFET is three, six and nine times higher for low, medium and high density SWCNTs compared to the devices with zero SWCNT. Similar significant increase can also be seen in the median value of the $I_{on}/I_{off}$ and $I_{on}$ with increasing SWCNT density (Figure 6(b), and 6(c)). For the devices with zero SWCNT electrodes, the median value of $I_{on}/I_{off}$ and $I_{on}$ are $1.5 \times 10^3$ and 0.6 µA, respectively. These values increased to 4.5 ×10³ (3 times) and 4.1 μA (7 times) for low, 2.0 ×10⁴ (17 times) and 8.3 μA (14 times) for medium, and 5.5×10⁴ (~40 times) and 11.82 μA (~20 times) for high SWCNT densities in the electrodes. From this study, it is clear that the density of SWCNT in the electrode, which controls the SWCNT/pentacene interfacial area, has significant impact on the performance of OFETs. Our study unequivocally show that, although a small number of SWCNTs in the electrodes can enhance the devices performance, the maximum performance were obtained using the most dense SWCNTs in the electrode.

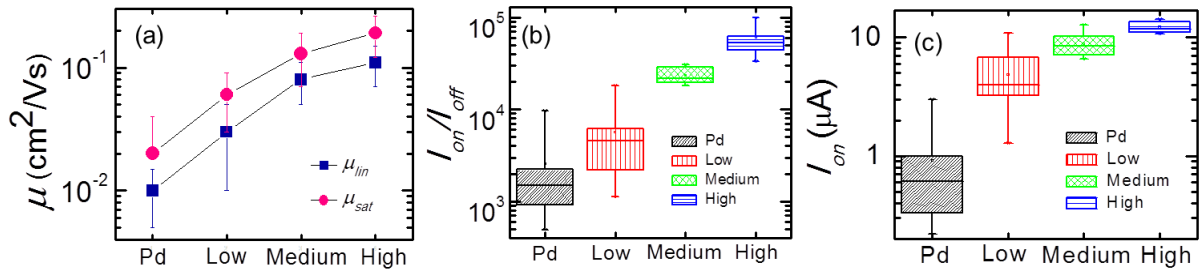

**Figure 6**. Summary of OFET devices performance from 40 devices. (a) Linear and saturation mobility. (b) On/off ratio and (c) On-current performance as a function of SWCNT density in the electrodes

**Table 1**: The saturation mobility ($\mu_{sat}$), linear mobility ($\mu_{lin}$), current on-off ratio ($I_{on}/I_{off}$) and on-current ($I_{on}$) for the devices with zero, low, medium, high density SWCNT in the electrodes.

|  |  | Pd (0 SWCNTs/μm) | Low density (1-5 SWCNTs/μm) | Medium density (10-15 SWCNTs/μm) | High density (25-30 SWCNTs/μm) |
|---|---|---|---|---|---|
| $\mu_{sat}$ | max | 0.05 cm²/Vs | 0.10 | 0.19 | 0.29 |
|  | average | (0.02 ± 0.02) | (0.06 ± 0.03) | (0.13 ± 0.06) | (0.19 ± 0.07) |
| $\mu_{lin}$ | max | 0.03 cm²/Vs | 0.06 | 0.13 | 0.19 |
|  | average | (0.01 ± 0.01) | (0.03 ± 0.02) | (0.08 ± 0.03) | (0.11 ± 0.04) |
| $I_{on}/I_{off}$ | max | 9.6x10³ | 1.8x10⁴ | 3.1x10⁴ | 1.1x10⁵ |
|  | median | 1.5x10³ | 4.5x10³ | 2.0x10⁴ | 5.5x10⁴ |
| $I_{on}$ | max | 3.3 μA | 10.8 | 12.8 | 14.2 |
|  | median | 0.6 μA | 4.1 | 8.3 | 11.8 |

The remarkable improvement in the OFET device performance with increasing the SWCNT density in the electrodes is due to increased interfacial area of SWCNT/pentacene interfaces. The current at an interface at a fixed bias voltage and temperature ($T$) can be approximated as $I \propto exp(-\varphi_b/KT)$, where $\varphi_b$ is the Schottky barrier between the metal/semiconductor interface and $K$ is the Boltzmann constant.[14] A decrease in $\varphi_b$ will result in an increase of current at the interface. It has been recently shown that the value of $\varphi_b$ at SWCNT/pentacene interface is ~ 0.16 eV which is much lower than the $\varphi_b$ at metal/pentacene interface (~0.35 to 0.85eV).[14] Figure 7 shows schematic diagrams of interfacial area for low and high density SWCNT electrodes. In the devices without any SWCNT, all the charge carriers are injected from Pd and pass through only Pd/pentacene interface. Since Pd has a larger barrier height compared to SWCNT, charge carriers need to overcome a larger injection barriers at the Pd/pentacene interface, which may reduce the number of injected charge carriers in the pentacene film and led to poor device performances. In contrast, when a small number of SWCNTs are anchored with Pd (low density SWCNT electrode) charge carriers are injected from both the SWCNT and Pd (Figure 7(a)). In this case, the injected charge carriers pass through a smaller barrier at SWCNT/pentacene and a larger barrier at Pd/pentacene. Since the charge carriers now have limited access of injection paths through SWCNT, the injection efficiency and device properties are improved. With increasing SWCNT densities, the carriers have larger SWCNT/Pentacene interfacial areas for more efficient charge injection through the lower barrier pathways (Figure 7(b)) and the device properties continues to improve resulting in higher device performance. It is important to note that, in our highest density electrodes there are 30 SWCNT/µm leaving an inter-nanotube separation of ~32 nm and we are unable to increase the density any further using DEP. If it will be possible to increase the density of SWCNT in the electrodes by any other technique, it can result in even more impressive device performance.

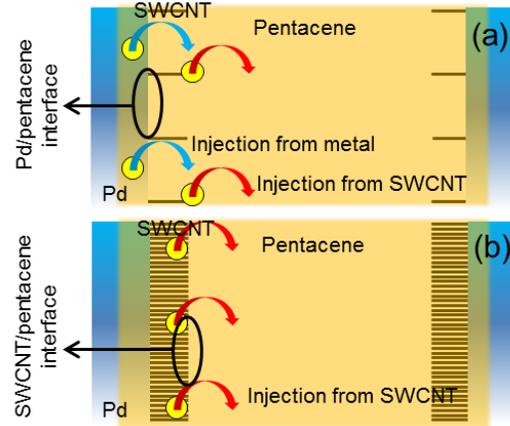

**Figure 7**. Schematic of interface with (a) low, and (b) high density SWCNT in the electrodes. The arrow indicates the charge carrier injection from the SWCNTs (red arrow) and Pd (blue arrow).

## 4. Conclusion

In summary, we investigated the performance of the pentacene transistors using aligned arrays SWCNT electrodes with various interfacial areas at the SWCNT/pentacene contact. From the electronic transport measurements of 40 devices, we showed that the OFET device performance such as mobility, current on-off ratio and on-current can be significantly improved with increasing interfacial area at the SWCNT/pentacene and best performance can be achieved by maximizing SWCNT/pentacene interfacial area. We attributed the improved device performance due to a lower barrier height at the SWCNT/pentacene interface compared to metal/pentacene interface.

**Acknowledgment**
This work is supported by U.S. National Science Foundation (NSF) under Grant ECCS 1102228